\documentclass[reprint, superscriptaddress, amsmath, amssymb,prl]{revtex4-2}
\usepackage{graphicx,float,dcolumn,bm,xcolor,nicefrac}
\usepackage[normalem]{ulem}
\definecolor{natcommblue}{RGB}{0, 104, 165}
\usepackage[colorlinks=true,citecolor=natcommblue,urlcolor=natcommblue,linkcolor=black]{hyperref}
\usepackage[pagewise]{lineno} 
\newcommand {\NNO}{NdNiO$_2$}
\newcommand {\NNOx}{NdNiO$_{2+x}$}
\newcommand {\NNOBM}{Nd$_3$Ni$_3$O$_7$}
\newcommand {\SLa}{$\mathbf{q}=\left(\nicefrac{1}{3},0\right)$}
\newcommand {\SLb}{$\mathbf{q}=\left(\nicefrac{1}{3},0,\nicefrac{1}{3}\right)$}
\newcommand {\supercell}{$3 \times 1 \times 3$}

\begin{document}
\title{Absence of $3a_0$ Charge Density Wave Order in the Infinite Layer Nickelates}
\author{C. T. Parzyck}
  \affiliation{Laboratory of Atomic and Solid State Physics, Department of Physics, Cornell University, Ithaca, NY 14853, USA}
\author{N. K. Gupta}
  \affiliation{Department of Physics and Astronomy, University of Waterloo, Waterloo ON N2L 3G1, Canada}
\author{Y. Wu}
  \affiliation{Laboratory of Atomic and Solid State Physics, Department of Physics, Cornell University, Ithaca, NY 14853, USA}
\author{V. Anil}
  \affiliation{Laboratory of Atomic and Solid State Physics, Department of Physics, Cornell University, Ithaca, NY 14853, USA}
\author{L. Bhatt}
  \affiliation{Department of Applied and Engineering Physics, Cornell University, Ithaca, NY 14853, USA}
\author{M. Bouliane}
  \affiliation{Department of Physics and Astronomy, University of Waterloo, Waterloo ON N2L 3G1, Canada}
\author{R. Gong}
  \affiliation{Department of Physics and Astronomy, University of Waterloo, Waterloo ON N2L 3G1, Canada}
\author{B. Z. Gregory}
  \affiliation{Laboratory of Atomic and Solid State Physics, Department of Physics, Cornell University, Ithaca, NY 14853, USA}
  \affiliation{Department of Materials Science and Engineering, Cornell University, Ithaca, NY 14853, USA}
\author{A. Luo}
  \affiliation{Department of Materials Science and Engineering, Cornell University, Ithaca, NY 14853, USA}
\author{R. Sutarto}
  \affiliation{Canadian Light Source, Saskatoon SK S7N 2V3, Canada}
\author{F. He}
  \affiliation{Canadian Light Source, Saskatoon SK S7N 2V3, Canada} 
\author{Y.-D. Chuang}
  \affiliation{Advanced Light Source, Lawrence Berkeley National Laboratory, 1 Cyclotron Road, MS 6-2100, Berkeley, California 94720, USA}  
\author{T. Zhou}
  \affiliation{Center for Nanoscale Materials, Argonne National Laboratory, Lemont, IL 60439, USA}
\author{G. Herranz}
  \affiliation{Institut de Ciència de Materials de Barcelona (ICMAB-CSIC), Campus UAB Bellaterra 08193, Spain}
\author{L. F. Kourkoutis}
  \affiliation{Department of Applied and Engineering Physics, Cornell University, Ithaca, NY 14853, USA}
  \affiliation{Kavli Institute at Cornell for Nanoscale Science, Cornell University, Ithaca, NY 14853, USA}
\author{A. Singer}
  \affiliation{Department of Materials Science and Engineering, Cornell University, Ithaca, NY 14853, USA} 
\author{D. G. Schlom}
  \affiliation{Department of Materials Science and Engineering, Cornell University, Ithaca, NY 14853, USA}
  \affiliation{Kavli Institute at Cornell for Nanoscale Science, Cornell University, Ithaca, NY 14853, USA}
  \affiliation{Leibniz-Institut f{\"u}r Kristallz{\"u}chtung, Max-Born-Stra{\ss}e 2, 12489 Berlin, Germany}  
\author{D. G. Hawthorn}
  \affiliation{Department of Physics and Astronomy, University of Waterloo, Waterloo ON N2L 3G1, Canada}
\author{K. M. Shen}
  \affiliation{Laboratory of Atomic and Solid State Physics, Department of Physics, Cornell University, Ithaca, NY 14853, USA}
  \affiliation{Institut de Ciència de Materials de Barcelona (ICMAB-CSIC), Campus UAB Bellaterra 08193, Spain}
  \affiliation{Kavli Institute at Cornell for Nanoscale Science, Cornell University, Ithaca, NY 14853, USA}

\begin{abstract}
A hallmark of many unconventional superconductors is the presence of many-body interactions which give rise to broken symmetry states intertwined with superconductivity. Recent resonant soft x-ray scattering experiments report commensurate $3a_{0}$ charge density wave order in the infinite layer nickelates, which has important implications regarding the universal interplay between charge order and superconductivity in both the cuprates and nickelates. Here, we present x-ray scattering and spectroscopy measurements on a series of {\NNOx} samples which reveal that the signatures of charge density wave order are absent in fully reduced, single-phase {\NNO}. The $3a_0$ superlattice peak instead originates from a partially reduced impurity phase where excess apical oxygens form ordered rows with 3 unit cell periodicity. The absence of any observable charge density wave order in {\NNO} highlights a crucial difference between the phase diagrams of the cuprate and nickelate superconductors. 
\end{abstract}
\maketitle

The discovery of superconductivity in the infinite-layer nickelates \cite{Li2019b}, and its analogy to its cuprate antecedents, offers a unique opportunity to better understand the key ingredients for high-temperature superconductivity. While the nickelates and cuprates share many commonalities, including a broadly similar crystal and electronic structure \cite{Li2019b}, strong correlations, antiferromagnetic excitations \cite{Lu2021,Fowlie2022}, and a superconducting dome \cite{Li2020a,Zeng2020a,Osada2021}, there are also many distinctions between the two families. These include the very different transition temperatures \cite{Li2019b,Li2020a,Zeng2020a}, the relative oxygen and $3d$ character of the doped holes \cite{Jiang2020,Goodge2021}, and the hybridization between the $3d$ and rare earth states \cite{Hepting2020}. One important apparent similarity between the two families is the report of charge density wave ordering in a variety of infinite layer nickelates by resonant soft x-ray scattering (RSXS) \cite{Rossi2022,Tam2022,Krieger2022,Ren2023}. This discovery, if correct, would suggest a ubiquitous interplay between charge order and superconductivity in the phase diagram of both the cuprate and nickelates, with important implications for a universal theory of high-temperature superconductivity. 

Nevertheless, there are clear distinctions between the charge order reported in the cuprates and the nickelates. In the cuprates, the wavevector is typically incommensurate and strongly doping dependent \cite{Blanco-Canosa2014,Fink2011,Hucker2011}, while in {\NNO} and PrNiO$_2$ it is locked to {\SLa} \cite{Tam2022,Krieger2022,Ren2023}.  Additionally, charge ordering in the cuprates is strongly temperature dependent \cite{Tranquada1995,Ghiringhelli2012,Comin2014,Blanco-Canosa2013,Fink2011,Hucker2011} whereas reports in nickelates exhibit a weak temperature dependence with no clear transition or onset \cite{Rossi2022,Krieger2022,Ren2023}. To understand better the nature of the putative charge ordering in the infinite-layer nickelates, we have investigated the {\SLa} superlattice peak in a series of samples with varying levels of reduction. We discover that the superlattice peak is entirely absent in fully reduced, single-phase NdNiO$_2$ samples, and instead arises from partially reduced impurity phases, NdNiO$_{2.33}$ (Nd$_3$Ni$_3$O$_7$) and/or NdNiO$_{2.67}$ (Nd$_3$Ni$_3$O$_8$), produced during the reduction process, where excess apical oxygen atoms form ordered rows with 3 unit cell periodicity. This reveals that charge ordering with $3a_{0}$ periodicity is not intrinsic to the infinite layer nickelates -- a discovery with important implications for our understanding of the phase diagram of the nickelates and its relationship to the cuprates.

\begin{figure*}[ht] \resizebox{17 cm}{!}{\includegraphics{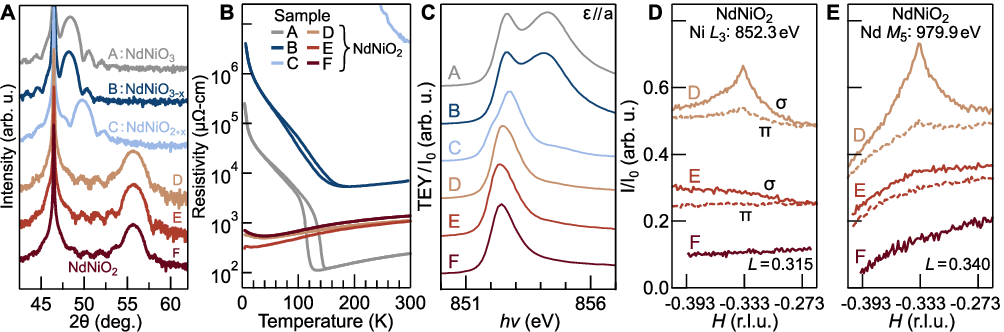}}
  \caption{\label{figure1} Characterization of perovskite NdNiO$_3$, oxygen deficient {\NNOx}, and infinite layer {\NNO}. (a) Cu $K_{\alpha}$ $\theta-2\theta$ x-ray diffraction measurements of a sequence of thin films ranging from NdNiO$_3$ (Sample A), to partially reduced {\NNOx} (Samples B \& C), to {\NNO} (Samples D-F). (b) Electrical transport measurements of the same series of samples. (c) Total electron yield XAS measurements of the Ni $L_3$ edge for this sample series. (d-e) Resonant soft x-ray scattering measurements of {\NNO} samples D, E, and F; rocking curves through the nominal charge order position of {$\mathbf{q}=\left(\nicefrac{-1}{3},0\right)$} at the Ni $L_3$ and Nd $M_5$ edges at $\sigma$ and $\pi$ polarizations, respectively.  Traces have been offset for clarity. }
\end{figure*}

To produce a sequence of {\NNO} samples with nominally identical Nd:Ni stoichiometry \cite{Li2021} and variable oxygen content, we have employed a combination of reactive oxide molecular beam epitaxy to synthesize the precursor perovskite and atomic hydrogen reduction to access the oxygen deficient phases.  A consistent series of perovskite films of NdNiO$_3$ (20 pseudocubic unit cells thick) with excellent crystallinity and sharp metal-insulator transitions were synthesized on SrTiO$_3$. Following synthesis, all films were capped by 2-3 unit cells of SrTiO$_3$ and exposed to a beam of $>50$\% atomic hydrogen produced by a thermal cracker \cite{Tschersich2008}. While reduction of the perovskite nickelates has typically been achieved using CaH$_2$ and NaH powder \cite{Hayward1999d,Li2019b,Zeng2020}, atomic H offers the benefit of independent control over the sample temperature and reducing environment, as well as fast reaction times ($<20$~min). While atomic hydrogen has prior been used as a reducing agent on binary oxides \cite{Bergh1965,Nishiyama2005}, this, to our knowledge, is its first application in the synthesis of complex oxides. Additional details about the reduction procedure are described in the Supplemental Information and will be detailed in a following publication. 

Using this approach, we have synthesized a series of samples ranging from the pristine parent perovskite NdNiO$_3$ (Sample A), to oxygen deficient intermediate phases {\NNOx} (Samples B and C), to infinite layer {\NNO} (Samples D-I), all of which were characterized by conventional x-ray diffraction (XRD) and transport measurements. The {\NNO} samples all appear highly crystalline and single phase by XRD, with low temperature resistivities of 350-700 $\mu\Omega -$cm, comparable to or lower than undoped films on SrTiO$_3$ and \cite{Zeng2020,Lee2020,Li2020a} and LSAT \cite{Lee2022}, as shown in Figures \ref{figure1}a-b and the Supplemental Information (for Samples G-I). 

X-ray absorption spectroscopy (XAS) measurements on all samples are shown in Figure \ref{figure1}c, and resonant soft x-ray scattering (RSXS) measurements on three representative infinite layer {\NNO} samples (D-F) are presented in Figures \ref{figure1}d and \ref{figure1}e. The XAS spectra of all three {\NNO} samples appear similar and match closely to the published measurements \cite{Rossi2021a,Hepting2020}, with a single peak at the Ni $L_3$ edge (852.4 eV) with no visible prepeak at the O $K$ edge. Despite the apparent similarity between these samples, only Sample D exhibits a superlattice peak at the putative charge order wavevector of {\SLa}. As in previous reports, this peak is observed on the Ni $L$ edge (Figure \ref{figure1}d) as well as the Nd (rare earth) $M$ edge (Figure \ref{figure1}e), and exhibits a strong polarization dependence, $I_{\sigma}/I_{\pi} \sim 4$ at $h\nu$ = 852.3 eV, consistent with prior measurements \cite{Ren2023,Tam2022}. The estimated in plane correlation length of $\xi = 12-20$~nm is also qualitatively similar to earlier reports \cite{Rossi2022,Tam2022,Krieger2022,Ren2023}. Finally, a weak dependence on the out-of-plane momentum transfer, $L$, is observed with a maximum at roughly $L$ = 0.31(2) reciprocal lattice units (r.l.u.), again consistent with earlier measurements (all momenta in this text are quoted in r.l.u. with reference to the {\NNO} lattice, $a=3.905$, $c=3.286$~\AA). In contrast, the superlattice peak was not observed in Samples E and F, which is notable since Sample E exhibited the lowest resistivity of all samples. Additional measurements on Samples G and H were measured at a separate beamline and exhibited an extremely weak {\SLa} feature at the sample center, while Sample I showed no superlattice features whatsoever. Further details of the measurements on Samples E-I are shown in the Supplemental Information. 

This discrepancy across nominally similar {\NNO} samples suggests that some unknown variable, potentially associated with sample disorder or oxygen nonstoichiometry, influences the presence of the $3a_0$ superlattice peak. One possibility is that charge density wave order is intrinsic to the infinite layer nickelates, but triggered by the presence of atomic-scale disorder within the lattice. Alternatively, the peak could originate from an impurity phase produced during the reduction process, which would imply that the charge order observed to date does not play a role in the phase diagram of the infinite layer nickelates.

\begin{figure}[t]
  \resizebox{1\columnwidth}{!}{\includegraphics{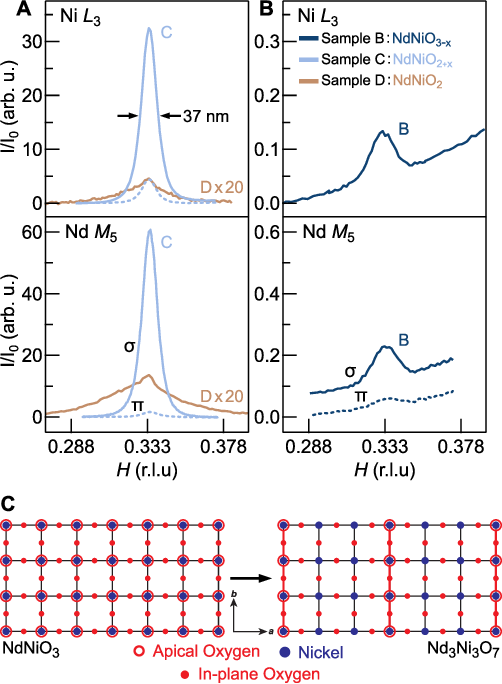}}
  \caption{\label{figure2} X-ray scattering measurements of oxygen deficient perovskite {\NNOx}, Samples B \& C. (a) $H$ scans, at constant $L$, at the Ni $L_3$ and Nd $M_5$ edges for sample C. Data for sample D, after removal of the fluorescent background and rescaling for visiblity, is included for reference. (b) Rocking curves through {\SLa} on lightly reduced Sample B, without background subtraction.  (c) Illustration of a potential intermediate phase structure for {\NNOBM} from a top-down view of the NiO$_2$ ab-plane where two thirds of the rows of apical oxygen atoms are removed, forming a $3a_0$ supercell along the a axis.}
\end{figure}

To distinguish between these scenarios, we have investigated partially reduced, oxygen deficient perovskite phases {\NNOx} (Samples B and C). Sample B was only lightly reduced, with its XRD nearly indistinguishable from the perovskite (Sample A), but with a substantially broadened metal-insulator transition and increased sample resistance (Figures \ref{figure1}a and b). Sample C was reduced further and shows an intermediate out-of-plane lattice constant of $c$=3.66~\AA~and highly insulating electrical transport as expected for a predominantly oxygen deficient perovskite phase \cite{Kawai2009}. The XAS spectra of Sample B exhibits the typical two-peak structure observed in perovskite nickelates on the Ni $L_{3}$ edge, \cite{Bisogni2016} as well as a strong prepeak at the O $K$ edge; Sample C is markedly different, with a sharp peak at 852.7 eV, shoulder at 852 eV, and weak secondary peak at 854.3 eV, indicative of an oxygen-deficient perovskite phase \cite{Abbate2002,Li2021b}.  

In Figure \ref{figure2}a we show RSXS measurements of Sample C which exhibits a peak at {\SLa}  virtually identical to the superlattice peak in Sample D and the published literature \cite{Rossi2022,Tam2022,Krieger2022,Ren2023} in nearly all respects: wavevector, energy dependence, polarization dependence, temperature dependence, and correlation length, with the exception that it is extremely intense (100-400 times stronger than is observed in Sample D). This peak also displays a strong $L$ dependence, with a maximum at $\sim 0.30$~ r.l.u ($d/3 = 3.65$~\AA).  The strong similarity between the superlattice features in Samples C and D suggests a common origin: ordered oxygen-deficient phases arising from an incomplete reduction process. In reduced nickelates,  excess apical oxygens can form ordered phases, such as the brownmillerite structure, La$_2$Ni$_2$O$_{5}$ \cite{Crespin1983a,Moriga1995,Alonso1995}, where the apical oxygen atoms form alternating rows. Other related structures also exist with different periodicities, such as Nd$_3$Ni$_3$O$_7$ \cite{Moriga1994a,Moriga2002,Wang2020}, Pr$_3$Ni$_3$O$_7$ \cite{Moriga1994a,Moriga2002}, La$_3$Ni$_3$O$_8$ \cite{Sayagues1994}, and (Pr,Ca)$_4$Ni$_4$O$_{11}$ \cite{Wu2023}. Recent \textit{in situ} x-ray diffraction studies indicate that the reduction pathway from bulk NdNiO$_3$ to {\NNO} occurs first via the formation of an intermediate phase {\NNOBM}, \cite{Wang2020}, where one third of the apical oxygen sites are occupied and are ordered into chains with with $3a_0$ periodicity (Figure \ref{figure2}c). This $3\times 1\times 3$ superstructure of the original pseudocubic unit cell would naturally give rise to a superlattice peak at {\SLb}. This suggests that Sample C is likely predominantly {\NNOBM}, and that an incomplete reduction of any {\NNO} samples would also leave traces of the {\NNOBM} phase behind (Sample D). 

\begin{figure}[!ht]
  \resizebox{1\columnwidth}{!}{\includegraphics{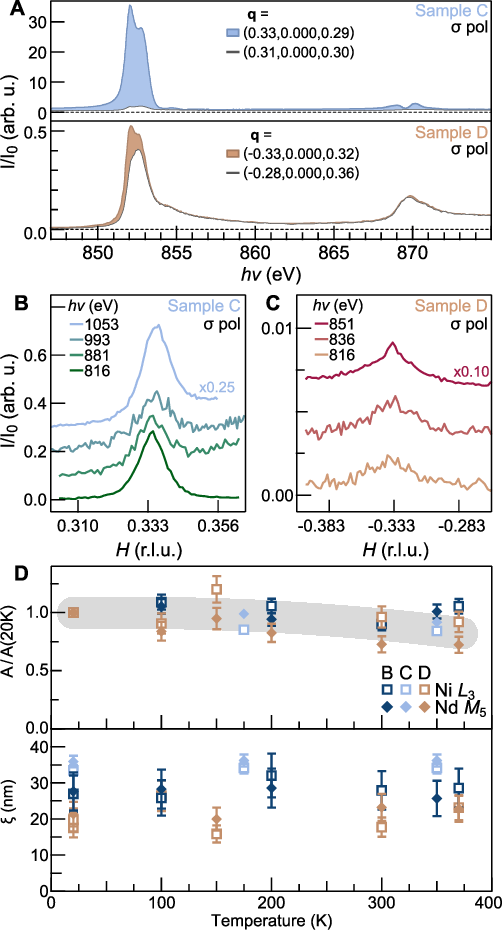}}
  \caption{\label{figure3} Energy and temperature dependence of the {\SLa} scattering peak. (a) Fixed wavevector resonant energy profiles both on and off of the scattering peak for samples C and D. The shaded region indicates intensity attributable to the resonant scattering above the fluorescent background (white). (b-c) Rocking curves through {\SLa} at a wide range of photon energies around the Ni $L$ and Nd $M$ resonances for samples C and D respectively; traces have been offset for clarity.  (d) Temperature dependence of the peak amplitude and correlation length ($\xi=2\pi/\Delta q$) at both edges for samples B, C, and D.  Error bars represent fitting uncertainties in extracted quantities; grey line is a guide to the eye.}
\end{figure}

To further investigate this hypothesis, we have also performed RSXS measurements on Sample B, which should only be in the initial stages of the conversion from NdNiO$_3$ into a brownmillerite-like {\NNOBM} phase. The XRD and XAS measurements of this sample (Figure \ref{figure1}a,c) are indistinguishable from pristine NdNiO$_3$ (Sample A). Nevertheless, RSXS measurements in Figure \ref{figure2}b also show the {\SLa} peak with identical characteristics ($L$, energy, and polarization dependence) and comparable intensity to Sample D. The fact that Sample B exhibits no obvious trace of the infinite layer phase, yet still exhibits a clear {\SLa} superlattice peak demonstrates that this feature does not originate from intrinsic charge ordering within the infinite layer phase itself. In fact, the {\SLa} peaks appear more reminiscent of Bragg peaks observed in resonant scattering from cuprates with oxygen ordering (e.g. ortho YBa$_2$Cu$_3$O$_{6+\delta}$), as opposed to intrinsic charge density wave order \cite{Hawthorn2011,Achkar2012}. There, the strong resonant enhancement on the Cu $L$ edge arises from the local oxygen environment strongly modifying the electronic structure of the Cu atoms, and a similar resonant enhancement at the Ni $L$ edge should likewise occur for oxygen ordering in the nickelates.

In Figure \ref{figure3}a, we show the energy dependence of the {\SLb} peak intensity in Samples C and D (shaded), together with the background fluorescence measured off the superlattice peak (white). Because the superlattice peak intensity is far stronger in Sample C, the relative strength of the background fluorescence in Sample D is much larger than in Sample C. Nevertheless, a two-peak resonance profile is apparent in both samples at the Ni $L_3$ edge with only a weak response at the $L_2$ edge, similar to prior measurements on NdNiO$_2$ \cite{Tam2022}. Assuming a structural origin arising from oxygen ordering, the superlattice peak should also be observable off-resonance. In Figures \ref{figure3}b and c, we show a series of scans across {\SLa} for Samples C and D, spanning a 235 eV range about the Ni and Nd resonances.  While the intensity is weaker than on resonance, the persistence of the peak strongly supports a structural origin. The temperature dependence of the scattering peak is shown for Samples B, C, and D in Figure \ref{figure3}d. Similar to prior reports \cite{Rossi2022,Krieger2022,Ren2023}, the peak shows a smooth decrease in intensity by 15-20\% between 20 and 370K, with no significant change in the correlation length. This weak dependence, absent of a transition, is quite similar to the temperature dependence of superlattice peaks resulting from the oxygen ordering in YBa$_2$Cu$_3$O$_{6+\delta}$ \cite{Strempfer2004,Achkar2012}, as opposed the more dramatic temperature dependence of the CDW order \cite{Ghiringhelli2012,Achkar2012}.

We also observe a strong dependence of the {\SLa} peak intensity in Sample G on the measurement location on the $10\times 10$~mm sample (Supplemental Information).  Due to the spread of the atomic hydrogen beam and thermal gradients, the edges of the sample are not as well reduced as the center -- correspondingly, we observe a clear superlattice peak near the edge of the sample which decreases in intensity approaching the sample center.  Spatially-resolved synchrotron hard x-ray diffraction measurements at the same locations confirm the prevalence of intermediate reduction products at the edge of Sample G, versus fully reduced NdNiO$_2$ near the center, thereby correlating the presence of the resonant feature with that of intermediate phases on the same sample.

\begin{figure*}[!ht]
  \resizebox{17 cm}{!}{\includegraphics{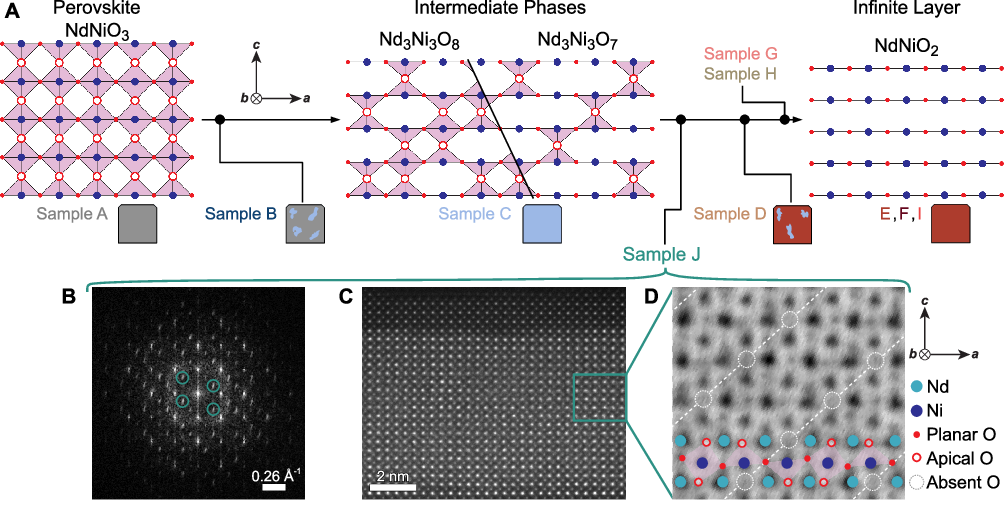}}
  \caption{\label{figure4} (a) Schematic of the reduction pathway from the perovskite NdNiO$_3$, through intermediate oxygen deficient phases {\NNOBM} or Nd$_3$Ni$_3$O$_8$, to the infinite layer NdNiO$_2$, and where the corresponding samples measured in this study lie on this pathway. (b-d) STEM images of a mixed phase sample (Sample J) containing Nd$_3$Ni$_3$O$_{7,8}$ intermediate phases. Fourier transform and corresponding HAADF image of a defective region are shown in (b) and (c), respectively. Positions of $3^{\textrm{rd}}$~order peaks in the Fourier transform are circled in green. (d) An ABF image showing two filled rows of apical oxygens, followed by one row of missing apical oxygen positions, corresponding to the schematic for Nd$_3$Ni$_3$O$_8$ in (a). In this image, the apical oxygen chains run into the page.}
\end{figure*}

Prior powder x-ray diffraction studies of bulk {\NNOBM} reveal that the excess apical oxygens form octahedrally coordinated chains that run perpendicular to the NiO$_2$ planes and which are separated by $3a_0$. Another potential arrangement is to have the apical oxygens form pyramidal chains which lie within the NiO$_2$ plane and are oriented along the $a$ axis, as shown in Figure \ref{figure4}a. In addition, Nd$_3$Ni$_3$O$_8$ (NdNiO$_{2.67}$), where two rows of the apical oxygens are occupied followed by a row of vacancies, would likewise generate the same superlattice peak at {\SLb}, although this phase has not been previously reported. Scanning transmission electron microscopy (STEM) measurements on a separate, partially reduced sample (Sample J), are detailed in Figure \ref{figure4}b-d.  The Fourier transform of a high angle annular dark field image (HAADF) reveals $\nicefrac{1}{3}$ order diffraction peaks corresponding to $3a_0$ ordering along both a and c. Additionally, annular bright field (ABF) data show regions with two filled apical oxygen chains alternating with a single vacant chain, consistent with the Nd$_3$Ni$_3$O$_8$ structure shown in Figure \ref{figure4}a, where the apical oxygen chains would be directed into the page. Sample J was prepared using the same conditions as Samples D-I but with a 4 u.c. SrTiO$_3$ cap. Although RSXS measurements were not performed on Sample J, XRD and transport measurements (Supplemental Information) indicate that this sample is less reduced than Sample D as both {\NNO} and intermediate phase peaks are visible in lab-based XRD. Since the difference in formation energies of the various oxygen ordered structures is small, it is possible that multiple compositions or structures of Nd$_3$Ni$_3$O$_{7,8}$ could exist within our series of samples, depending on factors such as the reduction conditions or epitaxial strain \cite{Gazquez2013}. A more detailed investigation of the precise structures of {\NNOBM} and/or Nd$_3$Ni$_3$O$_8$ in our thin films is currently the subject of further investigation. Nevertheless, the key point is that \emph{all} configurations of {\NNOBM} or Nd$_3$Ni$_3$O$_8$ will form a {\supercell} supercell that will generate the {\SLa} superlattice peak at the observed wavevector. Finally, a macroscopic sample would be expected to possess equal domains of chains along both in-plane directions, in which case the superlattice peaks would then be observable along both the $H$ and $K$ directions.


Our results suggest the scenario illustrated in Figure \ref{figure4}a. The reduction of NdNiO$_3$ into NdNiO$_2$ occurs via the production of an intermediate, partially reduced {\NNOBM} and/or Nd$_3$Ni$_3$O$_8$ phase, where excess apical oxygen atoms (or vacancies) are ordered into rows with $3a_0$ periodicity forming a {\supercell} supercell with a superlattice peak at the putative charge order wave vector of {\SLb} (Sample C). Because of its extremely strong intensity on resonance, this peak is detectable even if very small amounts of the partially reduced phase are present at levels nearly undetectable by conventional techniques (XRD, transport, and XAS; i.e. Samples B, D, and H). This picture is supported by a multitude of our observations including : 1) the superlattice peak is absent in the plurality of low resistivity, fully reduced {\NNO} samples (Samples E, F, and I), 2) the superlattice peak from Nd$_3$Ni$_3$O$_{7,8}$ (Sample C) exhibits nearly identical energy, polarization, and temperature dependence to {\NNO} Sample D and prior measurements on \textit{RE}NiO$_2$ \cite{Rossi2022,Tam2022,Krieger2022,Ren2023}, 3) the peak is commensurate to the lattice and observable well off resonance, indicating a strong structural component, 4) even a lightly reduced NdNiO$_3$ sample devoid of any detectable {\NNO} exhibits the superlattice peak (Sample B), 5) position-dependent measurements correlate the {\SLa} peak, observed by RSXS, with the presence of partially reduced phases, observed by hard x-ray diffraction at the same locations on the same sample (Sample G), and 6) the direct observation of excess apical oxygen ordering with $3a_0$ periodicity in a partially reduced sample by STEM (Sample J). 

The presence of an oxygen-ordered impurity phase also explains numerous discrepancies and unusual features in the published literature. The strong intensity on the rare-earth edge is surprising for a charge ordering scenario, since the coupling of the rare-earth $4f$ electrons to the Ni $3d$ electrons should be relatively weak and indirect, but should naturally occur for a structural Bragg peak where the Nd ions are displaced due to the oxygen ordering. In addition, prior reports claim the presence / absence of a superlattice peak in uncapped versus capped samples of NdNiO$_2$. Our measurements indicate that although the capping layer alone does not dictate the presence of the superlattice peak (all samples were capped), even small variations in reduction conditions can result in residual amounts of Nd$_3$Ni$_3$O$_{7,8}$. Finally, the doping dependence of the superlattice peak, which is strongest for the undoped parent compound and vanishes near $x$ = 0.20, can also be naturally explained by the presence of an impurity phase. Sr (hole) doping will increase the average targeted Ni valence of \textit{RE}$_{1-x}$Sr$_x$NiO$_2$, making the material easier to reduce. This will naturally diminish the amount of residual \textit{RE}$_3$Ni$_3$O$_{7,8}$ in the sample. This is supported by bulk studies \cite{Wang2020} which report that Sr doping substantially eases the reduction conditions of NdNiO$_3$ and facilitates the removal of oxygen from the system. Finally, although our study is limited to NdNiO$_{2+x}$, these findings should apply broadly to all the infinite layer nickelates, since Pr$_3$Ni$_3$O$_7$ and La$_3$Ni$_3$O$_8$ also form {\supercell} superstructures and are known to be reduction products of PrNiO$_3$ and LaNiO$_3$, respectively \cite{Moriga1994a,Moriga2002,Sayagues1994}. Although the superlattice peak in LaNiO$_2$ shares many overall similarities with those in (Nd,Pr)NiO$_2$, including its energy and temperature dependence, it does exhibit some subtle apparent differences, most notably a very slight displacement from the commensurate {\SLa} wavevector ($\Delta q \sim 0.01$~ r.l.u) \cite{Rossi2022} at $x$ = 0. Future experiments will be important for conclusively determining the origins of the superlattice peak in the other members of the infinite layer nickelates.

Through a multimodal investigation of a large sequence of samples with varying levels of reduction, we conclude that charge ordering with $3a_0$ periodicity is not intrinsic to the infinite layer nickelates. The superlattice peak previously identified as charge ordering at {\SLa} originates from the 3 unit cell ordering of excess apical oxygen ions in small amounts of brownmillerite-like inclusions of Nd$_3$Ni$_3$O$_7$ or Nd$_3$Ni$_3$O$_8$ produced during the reduction process. Topotactically reduced complex oxides present an exciting new frontier in quantum materials \cite{Li2019b,Kim2023}, but this work also highlights some of the materials challenges inherent in these systems. We demonstrate RSXS as a highly sensitive and powerful probe for investigating these reduced compounds which can detect even trace amounts of impurity phases. While the nickelate and cuprate phase diagrams show many similarities, including a superconducting dome and strong antiferromagnetic fluctuations on the underdoped side, this work establishes a clear distinction between the two families, namely that charge ordering does not appear to be directly relevant to the phase diagram of the nickelates. This finding should have important implications for understanding universal models of high-temperature superconductivity, and may help to explain some of the key differences between the two material families. 

\section*{Acknowledgements}
This work was primarily supported by the U.S. Department of Energy, Office of Basic Energy Sciences, under contract no. DE-SC0019414. This research used resources of the Advanced Light Source, a U.S. DOE Office of Science User Facility under Contract No. DE-AC02-05CH11231, as well as the Center for Nanoscale Materials and the Advanced Photon Source, both U.S. Department of Energy (DOE) Office of Science User Facilities operated for the DOE Office of Basic Energy Sciences by Argonne National Laboratory under Contract No. DE-AC02-06CH11357. Part of the research described in this paper was performed at the Canadian Light Source, a national research facility of the University of Saskatchewan, which is supported by the Canada Foundation for Innovation (CFI), the Natural Sciences and Engineering Research Council (NSERC), the National Research Council (NRC), the Canadian Institutes of Health Research (CIHR), the Government of Saskatchewan, and the University of Saskatchewan. Additional support for materials synthesis was provided by the Air Force Office of Scientific Research (Grant No. FA9550-21-1-0168), the National Science Foundation (No. DMR-2104427), and the Gordon and Betty Moore Foundation’s EPiQS Initiative through Grant Nos. GBMF3850 and GBMF9073. Substrate preparation was performed in part at the Cornell NanoScale Facility, a member of the National Nanotechnology Coordinated Infrastructure, which is supported by the NSF (Grant No. NNCI-2025233); the authors would like to thank Sean Palmer and Steven Button for their assistance in substrate preparation. STEM characterizations were performed at the Cornell Center for Materials Research Facilities supported by National Science Foundation (DMR-1719875). The microscopy work at Cornell was supported by the NSF PARADIM (DMR-2039380), with additional support from Cornell University, the Weill Institute and the Kavli Institute at Cornell. L.B and L.F.K. acknowledge support from Packard foundation. K.M.S. would like to acknowledge the hospitality of ICMAB-CSIC during his sabbatical. We would like to sincerely thank J. Fontcuberta, H.Y. Hwang, and W.S. Lee, and D.A. Muller for helpful discussions and insights. Author's note : During the preparation of this manuscript, we became aware of a report by transmission electron microscopy that observed the presence of ordered rows of excess apical oxygens with 3$a_0$ periodicity (arXiv:2306.10507).

\section*{Author contributions}
C.T.P. and K.M.S. conceived the research and designed the experiment. C.T.P, Y.W., and V.A. synthesized, reduced, and characterized the thin film samples, with input from D.G.S. and K.M.S. C.T.P, N.K.G., Y.W., V.A., M. B., R.G., R.S., Y.-D.C, D.G.H., and F.H. performed the resonant soft x-ray scattering measurements. L.B. and L.F.K. performed the scanning transmission microscopy experiments. B.G., A.L., T.Z., and A.S. performed the synchrotron hard x-ray diffraction experiments. The results were analyzed and interpreted by C.T.P., N.K.G., Y.W., V.A., G.H., D.G.H., and K.M.S. C.T.P. and K.M.S. wrote the manuscript, with input from all authors. 

\bibliography{Nickelates}
\end{document}


\title{Supplemental Information For: Absence of $3a_0$ Charge Density Wave Order in the Infinite Layer Nickelates}
\author{C. T. Parzyck}
  \affiliation{Laboratory of Atomic and Solid State Physics, Department of Physics, Cornell University, Ithaca, NY 14853, USA}
\author{N. K. Gupta}
  \affiliation{Department of Physics and Astronomy, University of Waterloo, Waterloo ON N2L 3G1, Canada}
\author{Y. Wu}
  \affiliation{Laboratory of Atomic and Solid State Physics, Department of Physics, Cornell University, Ithaca, NY 14853, USA}
\author{V. Anil}
  \affiliation{Laboratory of Atomic and Solid State Physics, Department of Physics, Cornell University, Ithaca, NY 14853, USA}
\author{L. Bhatt}
  \affiliation{Department of Applied and Engineering Physics, Cornell University, Ithaca, NY 14853, USA}
\author{M. Bouliane}
  \affiliation{Department of Physics and Astronomy, University of Waterloo, Waterloo ON N2L 3G1, Canada}
\author{R. Gong}
  \affiliation{Department of Physics and Astronomy, University of Waterloo, Waterloo ON N2L 3G1, Canada}
\author{B. Z. Gregory}
  \affiliation{Laboratory of Atomic and Solid State Physics, Department of Physics, Cornell University, Ithaca, NY 14853, USA}
  \affiliation{Department of Materials Science and Engineering, Cornell University, Ithaca, NY 14853, USA}
\author{A. Luo}
  \affiliation{Department of Materials Science and Engineering, Cornell University, Ithaca, NY 14853, USA}
\author{R. Sutarto}
  \affiliation{Canadian Light Source, Saskatoon SK S7N 2V3, Canada}
\author{F. He}
  \affiliation{Canadian Light Source, Saskatoon SK S7N 2V3, Canada}
\author{Y.-D. Chuang}
  \affiliation{Advanced Light Source, Lawrence Berkeley National Laboratory, 1 Cyclotron Road, MS 6-2100, Berkeley, California 94720, USA}
\author{T. Zhou}
  \affiliation{Center for Nanoscale Materials, Argonne National Laboratory, Lemont, IL 60439, USA}
\author{G. Herranz}
  \affiliation{Institut de Ciència de Materials de Barcelona (ICMAB-CSIC), Campus UAB Bellaterra 08193, Spain}
\author{L. F. Kourkoutis}
  \affiliation{Department of Applied and Engineering Physics, Cornell University, Ithaca, NY 14853, USA}
  \affiliation{Kavli Institute at Cornell for Nanoscale Science, Cornell University, Ithaca, NY 14853, USA}
\author{A. Singer}
  \affiliation{Department of Materials Science and Engineering, Cornell University, Ithaca, NY 14853, USA}
\author{D. G. Schlom}
  \affiliation{Department of Materials Science and Engineering, Cornell University, Ithaca, NY 14853, USA}
  \affiliation{Kavli Institute at Cornell for Nanoscale Science, Cornell University, Ithaca, NY 14853, USA}
  \affiliation{Leibniz-Institut f{\"u}r Kristallz{\"u}chtung, Max-Born-Stra{\ss}e 2, 12489 Berlin, Germany}
\author{D. G. Hawthorn}
  \affiliation{Department of Physics and Astronomy, University of Waterloo, Waterloo ON N2L 3G1, Canada}
\author{K. M. Shen}
  \affiliation{Laboratory of Atomic and Solid State Physics, Department of Physics, Cornell University, Ithaca, NY 14853, USA}
  \affiliation{Institut de Ciència de Materials de Barcelona (ICMAB-CSIC), Campus UAB Bellaterra 08193, Spain}
  \affiliation{Kavli Institute at Cornell for Nanoscale Science, Cornell University, Ithaca, NY 14853, USA}

\maketitle
\tableofcontents
\newpage

\newpage
\section{Experimental Methods}
Thin films of NdNiO$_3$ were grown on SrTiO$_3$ substrates using reactive-oxide molecular beam epitaxy in a Veeco GEN10 system using elemental beams of Nd (Alfa/AESAR, 99.9\%) and Ni (Alfa/AESAR, 99.995\%).  Substrates were etched to prepare a TiO$_2$ terminated surface and annealed prior to growth at 650 $^{\circ}$C until a clear RHEED pattern was observed. Growths were performed at substrate temperatures between 480 and 500 $^{\circ}$C in background pressures between 2 and $6\times10^{-6}$ torr of 80\% distilled ozone. Initial flux calibration was performed by monitoring RHEED oscillations during the growths of binary oxides Nd$_2$O$_3$ on (ZrO$_2$)$_{0.905}$(Y$_2$O$_3$)$_{0.095}$ (111) and NiO on MgO (100) using the parameters outlined in Ref. \onlinecite{Sun2022}.  Further stoichiometric optimization was performed by minimizing the NdNiO$_3$ (002)$_{\textrm{pc}}$ plane spacing \cite{Li2021}.  Following growth of the nickelate layer, a SrTiO$_3$ capping layer was grown at 500 $^{\circ}$C in a background pressure of $2\times10^{-6}$~torr, following calibration of the Sr and Ti sources via monitoring of RHEED oscillations.  Samples were reduced using a beam of atomic hydrogen produced by a thermal source \cite{Tschersich1998,Tschersich2008} where molecular H$_2$ is passed through a heated tungsten capillary ($>1650$~$^{\circ}$C) where it disassociates into individual atoms \cite{Zheng2006} before interacting with the sample. Reductions were performed in an ultra-high vacuum chamber ($P_{\textrm{base}}<1\times 10^{-10}$~torr) located on the same vacuum manifold as the MBE growth system at temperatures between 250 and 310 C and hydrogen fluxes ranging between 1.8 and $2.7 \times 10^{15}$~at/cm$^2$/sec. Typical reductions involved between 12 and 15 minutes of exposure to the atomic hydrogen beam to produce samples appearing fully reduced by lab-based XRD (Samples D-I).  Structural quality of both the perovskite and reduced samples were determined using Cu K$\alpha_1$ x-ray diffraction measurements performed on a PANalytical Empyrean X-ray diffractometer. Electrical transport measurements were performed using both a custom built LHe cooled four point probe measurement station as well as a Quantum Design physical property measurement system (base temperature of 2 K) with finite size factors accounted for using the methods of Ref. \onlinecite{Miccoli2015}. Contacts were prepared by either application of a dot of indium metal underneath a gold contact pin, or by ultrasonic aluminum wire bonding.

Soft x-ray scattering measurements were performed at the REIXS beamline of the Canadian Light Source on a 4-circle diffractometer in an ultrahigh-vacuum chamber ($P<5\times10^{-10}$~ torr). The nominal photon flux and energy resolution were $I_{0}=5\times10^{11}$ photons/sec and $\Delta E/E \sim 2\times10^{-4}$, respectively. The incoming x-ray polarization was selected to be in either the $\sigma$ ($\upvarepsilon \perp$~scattering plane) or $\pi$ ($\upvarepsilon \parallel$  scattering plane) configuration with the polarization of the scattered x-rays unmeasured.  Measurements were conducted at photon energies between 815 and 1050 eV with either a micro-channel plate (MCP) or a silicon drift detector (SDD) with an angular acceptance of $0.9^{\circ}$.  The SDD is an energy-resolved detector, with a resolution larger than 50 eV, which allows for removal of the substantial oxygen fluorescence background produced by the SrTiO$_3$ substrate.  Precise alignment to the film crystallographic axes was achieved by detecting the $(001)$, $(101)$, and $(\overline{1}01)$ Bragg peaks of the SrTiO$_3$ substrate with an energy of 2.5 keV.  Additional measurements detailed in section IV of this supplemental were performed at beamline 8.0.1 of the Advanced Light Source.  For these measurements a fixed $\pi$ polarization geometry was used and the nominal photon flux and energy resolution were $\sim 10^{13}$ photons/sec and $\Delta E/E \sim 7\times10^{-4}$, respectively.  The fluorescence yield and scattering signal were recorded using a GaAsP photodiode (with an Al window to block visible light and photoelectrons) mounted at 100 mm from the sample with an acceptance angle of 3$^{\circ}$.  To consistently calibrate the photon energies between different beamlines a perovskite nickelate sample (Sample A) was measured at both endstations as an energy reference.  The energy scale was then defined with respect to the NdNiO$_3$~Ni $L_3$ peaks at 852.6 and 854.3 eV and O $K$ prepeak at 527.9 eV to align with prior literature measurements \cite{Kim2023,Bisogni2016,Hepting2020,Rossi2022}. Here, and in the main text, all momenta are quoted in reciprocal lattice units (r.l.u.) relative to the {\NNO} lattice, $a=3.905$~\AA, $c=3.286$~\AA.

 Scanning x-ray micro-diffraction measurements were conducted at beamline 26-ID of the Advanced Photon Source at Argonne National Laboratory. We used a liquid-nitrogen cooled Si(111) double crystal monochromator to achieve an energy resolution of $\Delta E / E = 10^{-4}$ at a photon energy of 10.0 keV and an area detector with each pixel extending 0.00075 \AA$^{-1}$~ along the Ewald sphere. We reduced the collimated x-ray beam spot to 100 $\mu$m (FWHM) with slits, yielding a flux of $\sim 1010$~photons/s. Rocking the sample by $4^{\circ}$ in 41 steps and repeatedly measuring a spatial line scan at each angle, we obtained a 3D reciprocal space map (RSM) every 100 $\mu$m along a 4 mm line. Projections of the maps onto the L-axis of the substrate/film are shown in Figure \ref{fig:microXRD}(b). All micro-diffraction data presented in this work was obtained by rastering vertically along the sample, perpendicular to the scattering plane, minimizing the beam footprint and ensuring the spatial resolution of the line scans matches the beam width of 100 $\mu$m. 

STEM characterization was performed on a cross-sectional lamella prepared with the standard focused ion beam (FIB) lift-out procedure using a ThermoFisher Helios G4 UX FIB. HAADF- and ABF-STEM imaging were performed on Cs-corrected Thermo FisherScientific Spectra at 300 kV and 30 mrad probe convergence semi-angle. For high-precision structural measurements, a series of 40 rapid-frame images were acquired and subsequently realigned and averaged by a method of rigid registration optimized to prevent lattice hops \cite{Savitzky2018} resulting in a high signal-to-noise ratio, high fidelity image of the atomic lattice.

\clearpage
\section{XAS Analysis and Intermediate Phase Structures}
\begin{figure}[ht]
  \resizebox{14 cm}{!}{\includegraphics{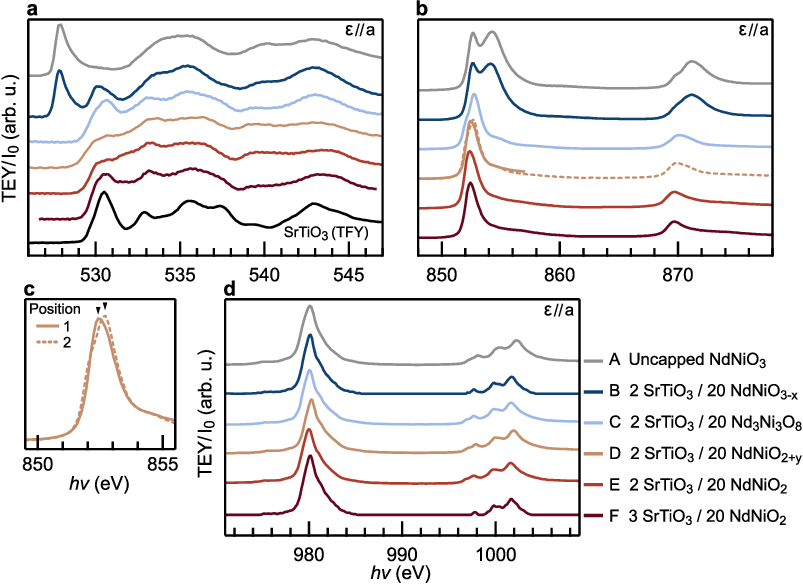}}
  \caption{\label{fig:morexas} Additional XAS measurements.  (a) Total electron yield (TEY) Measurements at the oxygen K-edge for all samples presented in the main text.  Total fluorescence yield (TFY) spectrum of a bare SrTiO$_3$ substrate is provided for reference. (b) Extended range spectrum at the Nickel $L_2$ and $L_3$ edges. (c) Zoom in on Ni $L_3$ XAS for sample D, showing spectra obtained at two positions near the sample center. (d) TEY spectra at the Neodymium $M_4$ and $M_5$ edges.  All spectra were collected at normal incidence at temperatures of either 20 K or 78 K; traces have been rescaled and offset for clarity.}
\end{figure}

Here we present additional x-ray absorption spectroscopy (XAS) measurements on the samples from the main text.  Larger range XAS measurements at the Oxygen $K$, Nickel $L$, and Neodymium $M$ edges are shown in Figure \ref{fig:morexas}.  All spectra were collected at normal incidence with either a silicon drift detector or photodiode at either 20 or 78 K.  Measurements were performed at both beamline 8.0.1 of the Advanced Light Source and the REIXS beamline of the Canadian Light Source and a common NdNiO$_3$ samples (Sample A) was measured at both beamlines as an energy reference.  At the O $K$ edge, samples A and B show a strong pre-peak associated with the O $2p$ - Ni $3d$ hybridization \cite{DeGroot1989,Palina2017} typical of the the perovskite nickelates.  Conversely, samples C-F show no evidence of a prepeak feature at $\sim 528$~eV, indicative of full conversion of the original perovskite to other phases \cite{Goodge2021}.  At the Ni $L$ edge, a split-peak structure at the nickel $L_3$ edge (851-855 eV) is observed in both samples A and B, as is expected for the perovskite nickelate, and no additional peaks associated with the formation of NiO are observable, indicating complete oxidation of the as-grown perovskite samples.  The infinite layer samples (D-F) show the typical single peak structure reported in the literature \cite{Chen2022,Rossi2021a}.  Two spectra from different positions near the sample center are reported for Sample D in Figure \ref{fig:morexas}(c). The two positions show slightly different XAS profiles at the L$_3$ edge with a peaks at 852.4 eV and 852.7 eV at positions 1 and 2, respectively.  The resonant feature at {\SLa} was observed at both positions on the sample with similar correlation lengths ($\xi_2\sim\xi_1$) but different amplitudes ($A_2 \sim 3 A_1$).  Finally, the Neodymium $M$ edges, shown in Figure \ref{fig:morexas}(d), show little variation across the sample series.

In addition to the TEY measurements presented in Figure \ref{fig:morexas}, which were performed on all samples, Partial Fluorescence Yield (PFY) measurements at the Ni $L$ edge were also measured on samples B, C and F using a silicon drift detector \cite{Eisebitt1993,Achkar2011}.  Selecting fluorescence from Ni $L$ edge emission, and eliminating emission from oxygen or titanium, enables measurement of emission from the sample without contributions from the SrTiO$_3$ (STO) substrate or capping layer.  These measurements were performed with both $\sigma$ and $\pi$ polarized light for a range of incident angles spanning grazing incidence ($\theta = 10^\circ$) to normal incidence ($\theta = 90^\circ$), such that $\pi$ polarized light at grazing incidence probes with the electric field approximately parallel to the $c$-axis of the sample.

For a thin film of thickness $d_{NNO}$, the intensity of the Ni $L$ edge partial fluorescence yield is given by
\begin{equation}
\begin{aligned}
I_{NiL,PFY} & \propto \frac{\mu_{NiL,i}}{\sin\alpha}\left(\int_{0}^{d_{NNO}}e^{-\left(\frac{\mu_{tot,i}}{\sin\alpha}+\frac{\mu_{tot,f}}{\sin\beta}\right)z}dz\right)\\
&=\frac{\mu_{NiL,i}}{\sin\alpha}\left(\frac{1}{ \frac{\mu_{tot,i}}{\sin\alpha}+\frac{\mu_{tot,f}}{\sin\beta} }\right)\left(1-e^{-\left(\frac{\mu_{tot,i}}{\sin\alpha}+\frac{\mu_{tot,f}}{\sin\beta}\right)d_{NNO}}\right)\\
&=\frac{\mu_{NiL,i}d_{NNO}}{\sin\alpha}\left(1-\frac{1}{2}\left(\frac{\mu_{tot,i}}{\sin\alpha}+\frac{\mu_{tot,f}}{\sin\beta}\right)d_{NNO} + \frac{1}{3}\left(\frac{\mu_{tot,i}}{\sin\alpha}+\frac{\mu_{tot,f}}{\sin\beta}\right)^2d_{NNO}^2 + ...\right)
\end{aligned}
\label{eq:PFY}
\end{equation}
where $\mu_{tot,i}$($\mu_{tot,f}$) is the total linear x-ray absorption co-efficient of the NNO$_x$ layer at the incident(emitted) photon energy and $\mu_{NiL,i}$ is the contribution to the Ni $L$ absorption coefficient of NNO$_x$ layer from excitation of $2p$ core electrons into unoccupied states.  In our samples, the thickness of the STO capping layer $d_{STO} \sim 7 - 12 \mathrm{\AA}$ is thin relative to $1/\mu_{STO} \sim 2800 \mathrm{\AA}$. Accordingly, the STO layer provides negligible absorption of the incident and scattered beam.  As shown in equation~\ref{eq:PFY}, the intensity of the Ni L edge PFY will be proportional to $\mu_{NiL,i}$, but will be subject to a self-absorption correction depending on the values of $\mu_{tot,i}$, $\mu_{tot,f}$, $\alpha$, $\beta$ and $d_{NNO}$.

\begin{figure}[ht]
  \begin{center}
    \includegraphics[width=6.0in]{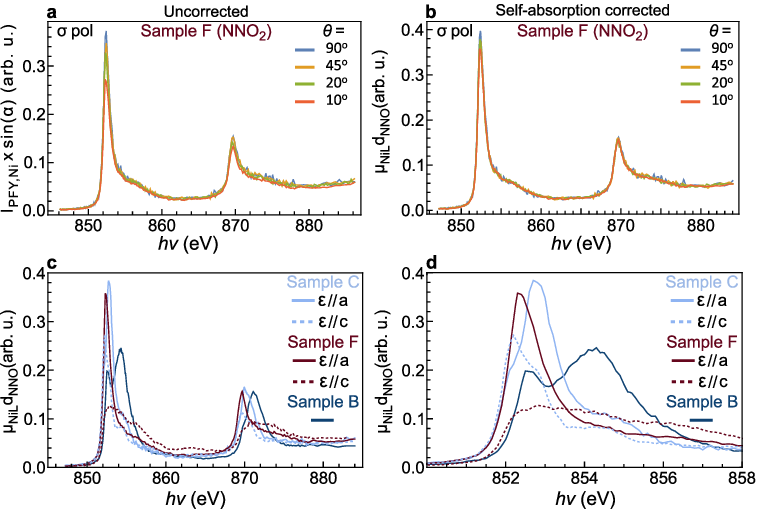}
    \caption{\label{fig:figSNiPFYa} Ni $L$ edge Partial Fluorescence yield measurements of the x-ray absorption. (a) The Ni $L$ edge partial fluorescence of sample F as a function of incidence angle, $\alpha$ with $\sigma$ polarized light ($E \parallel a$). Measurements are shown normalized to the incident beam intensity, $I_0$ and $\sin \alpha$.  (b) $\mu_{NiL,i}$ determined from the data in a by applying a self-absorption correction.  (c) and (d) $\mu_{NiL,i}$ as a function of energy for sample C and F for both $E \parallel a$ and $E \parallel c$.  Data for $E \parallel c$ was deduced from measurements at grazing incidence ($\alpha = 10^\circ$) with $\pi$ polarized light.}
  \end{center}
\end{figure}

In figure~\ref{fig:figSNiPFYa}, the raw PFY data for sample F is shown as a function of angle of incidence, $\alpha$, for $\sigma$ incident light.  Here the data is normalized only to $\sin \alpha$ and the incident beam intensity, $I_0$.  The good agreement between measurements at different $\alpha$ values above the $L_3$ edge indicates that self absorption corrections are not too large ($> 10\%$).  However, at the peak of the $L_3$ edge at grazing incidence, self absorption can reduce the peak intensity by $\sim 25\%$, for our 20 u.c. thick samples.  This self absorption effect can be partially corrected by scaling the PFY data to tabulated values of the absorption co-efficient above the Ni $L$ edge \cite{Chantler2000} and inverting equation~\ref{eq:PFY}. This procedure gives $\mu_{NiL,i}d_{NNO}$ shown in figure~\ref{fig:figSNiPFYa}b, showing good agreement between measurements at different values of $\alpha$.  
Applying this procedure to sample B, C and F using $\sigma$ and $\pi$ polarized light at grazing incidence provides a good comparison of the energy and polarization dependence of the XAS between samples, notably without any free parameters or the need to otherwise scale or offset measurements taken on different samples. As shown in figure~\ref{fig:figSNiPFYa} c and d, the NdNiO$_2$ sample (Sample F) shows dichroism consistent with literature, with a large peak for $E \parallel a$ at 851.2 eV that is diminished with $E \parallel c$, consistent with Ni$^{1+}$ and holes in $d_{x^2-y^2}$ orbitals.

In contrast, sample C exhibits a peak at slightly higher energy (852.8 eV) than sample F for $E \parallel a$. Moreover, with $E \parallel c$, sample C exhibits a peak at lower energy (852.2 eV) than the peaks in both samples C and F with $E \parallel a$.  Moreover, the energy dependence and dichroism, contrasts with that of perovskite Ni$^{3+}$ (sample B).  This is suggestive of at least some concentration of Ni$^{2+}$ sites in sample C,\cite{Haverkort2010} with holes in both  $d_{x^2-y^2}$ and $d_{3z^2-r^2}$ orbitals. Unlike Ni$^{2+}$ in NiO, dichroism indicates a crystal field with symmetry lower than O$_{h}$, such as Ni residing in pyramidal sites, with $d_{3z^2-r^2}$ orbitals having a lower energy than the $d_{x^2-y^2}$ states.  

\begin{figure}[!ht]
  \resizebox{17 cm}{!}{\includegraphics{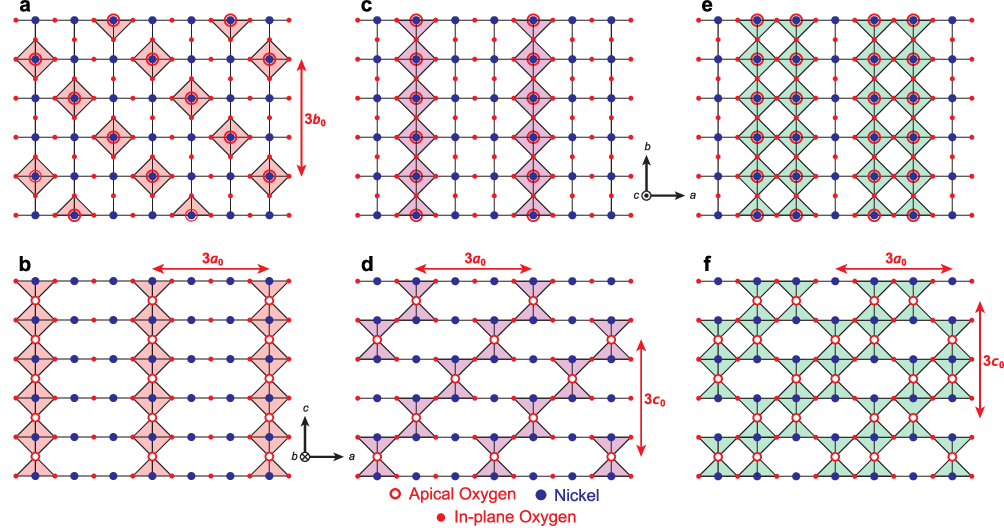}}
  \caption{\label{fig:cartoon}Proposed structures for they oxygen ordering in NdNiO$_{2+x}$. (a-b) The proposed structure of Nd$_3$Ni$_3$O$_7$, from Moriga et al. based on x-ray powder refinement, with octahedral coordination of the nickel ions. (c-d) A proposed structure for Nd$_3$Ni$_3$O$_7$ consisting of alternating square-planar and pyrimidal coordination of the nickel ions. (e-f) A proposed structure for Nd$_3$Ni$_3$O$_8$  consisting of alternating pyrimidal and octahedral coordination of the nickel ions.}
\end{figure}

In Figure \ref{fig:cartoon} we display three proposed structures for the oxygen ordering in intermediate phase nickelates.  Figure \ref{fig:cartoon}(a-b) reproduces the structure for (Nd,Pr)$_3$Ni$_3$O$_7$ proposed based on bulk powder x-ray refinement in Ref. \cite{Moriga1994a} with the remaining oxygen atoms forming octahedral chains along one axis.  XAS measurements, detailed in Figure \ref{fig:figSNiPFYa}, on sample C suggest a lower symmetry configuration of the nickel ions than the O$_{h}$ configuration proposed by Moriga et al. \cite{Moriga1994a}.  As such we propose an alternate structure for Nd$_3$Ni$_3$O$_7$ / SrTiO$_3$ with alternating pyramidal and planar coordination of the nickel ions, and chains of missing apical oxygen ions running in the basal plane.  This structure, depicted in panel (c-d), can be naturally extended via the addition of one additional oxygen ion per formula unit to a structure of Nd$_3$Ni$_3$O$_8$ which is consistent with the STEM results on Sample J presented in the main text, as shown in Figure \ref{fig:cartoon}(e-f).  Prior studies of oxygen deficient brownmillerite-like structures, (La,Sr)$_2$Co$_2$O$_5$, indicate that the ordering of oxygen vacancies is sensitive to sample orientation and strain \cite{Gazquez2013}. Furthermore, bulk electron diffraction studies of the reduction products of LaNiO$_3$ indicate the presence of several orderings, even in the same sample, including La$_2$Ni$_2$O$_5$, La$_4$Ni$_4$O$_{11}$, and La$_3$Ni$_3$O$_8$ \cite{Gonzalez1989,Sayagues1994}.  As such, it is plausible that any or all of these structures may occur in a given thin film sample, with their presence dictated by details of the sample (e.g. the presence or absence of defects in the perovskite precursor) or the reduction process itself (reducing agent, time, temperature). Detailed macro- and micro-structural analysis of the factors contributing to oxygen ordering in the infinite layer nickelates, and topotactically reduced oxides generally, will likely play an important role in both stabilizing and understanding novel reduced oxides.

\section{Additional data and measurements on samples A-F}

\subsection{Temperature Dependence}
\begin{figure}[ht]
  \resizebox{17 cm}{!}{
  \includegraphics{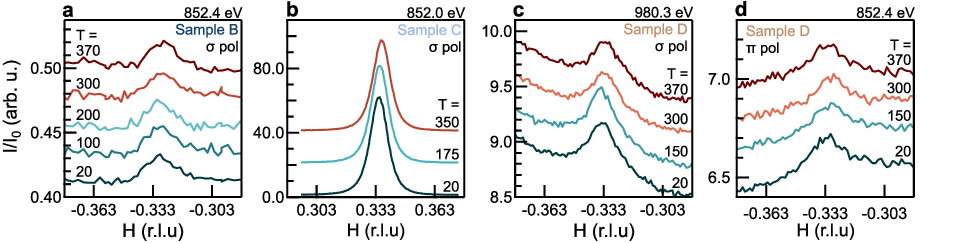}}
  \caption{\label{fig:temps2} Additional temperature dependent measurements of the resonant feature for samples B, C, and D of the main text. (a) Measurements of the lightly reduced perovskite sample, B, at the Ni $L_3$ edge. (b) Measurements of the intermediate oxygen ordered phase sample, C, at the Ni $L_3$ edge. (c-d) Measurements of the predominately infinite layer sample, D, at the Nd $M_5$ and Ni $L_3$ edges, respectively.}
\end{figure}
A selected set of the raw data from the temperature dependent measurements depicted Figure 3 of the main text are presented in Figure \ref{fig:temps2}.  As can be seen in the raw data, there is little to no observable variation in the peak characteristics in any of samples B, C, or D as a function of temperature, which is consistent with the fitting results presented in the main text -- only a weak dependence in the intensity and no observable trend in the correlation length.  We note that the fitted correlation correlation lengths, $\xi = 2\pi/\Delta q$, observed here are somewhat larger then those reported previously by a factor of 3-5 \cite{Krieger2022,Tam2022,Rossi2022,Ren2023}.  Finally, we observe that in each temperature series that there is a slight shift in the peak position to higher $q$ as the samples are warmed from 20 to 370~K.  In sample C the shift is measured to be 0.0012 r.l.u. (0.3\%) relative to SrTiO$_3$ in-plane lattice parameter. This shift is consistent with thermal expansion of the SrTiO$_3$ substrate as measured in Ref. \onlinecite{Loetzsch2010}. The increase in the pseudocubic lattice parameter from $\sim 3.899$~\AA at 20~K to 3.905~\AA at 300~K shifts reciprocal lattice to higher $q$ by $\sim 0.2$\%, in good agreeance with our observations.

\subsection{Out-of-plane Momentum Transfer}
\begin{figure}[ht]
  \resizebox{14.3 cm}{!}{
  \includegraphics{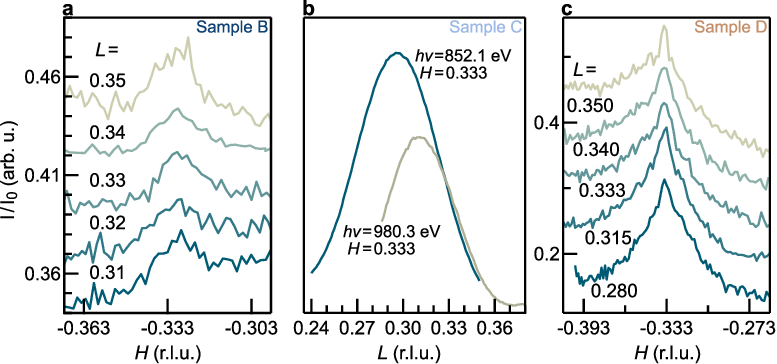}}
  \caption{\label{fig:ldep} $L$ dependence of the resonant feature. (a) Rocking curves through (-0.333, 0, L) for different values of $L$ on sample B from the main text.  $L$ scans at constant H for sample C at the Nd/Ni resonances, after application of an angle dependent absorption correction.  (c) Rocking curves through $(-0.333, 0, L)$ at different $L$ from sample D.  Traces have been offset for clarity}
\end{figure}
Dependence on the out-of-plane momentum transfer, $L$, for samples B, C, and D is reported in \hyperref[fig:ldep]{Figure \ref*{fig:ldep}}; in all cases $L$ is reported relative to the infinite layer out-of-plane lattice parameter, $c=3.286$~\AA.  In sample B, little change is observed as a function of of $L$ over the range measured, indicating short correlation lengths out of the plane of the ordering.  In contrast, in the film composed predominately of the ordered oxygen vacancy phase shows a relatively sharp $L$ dependence, though with a larger correlation length ($\xi_{(0,0,L)}=$4-6 nm) than in-plane ($\xi_{(H,0,0)}=37$~nm). This is expected given the out-of-plane coherence is limited by the film thickness of $\sim 7$~nm.  At 852.1 eV, the curve is peaked around L = 0.30 r.l.u., which gives a plane spacing of $d/3 = 3.65$~\AA ~which is in agreement with the film c-axis lattice constant, $c=3.659$~\AA, measured by XRD (Main text Figure 1).  Note, the data shown in Figure \ref*{fig:ldep}(b) is corrected for angle-dependent absorption of the incident and scattered beam by dividing the scattered intensity by 
\begin{equation}
\frac{1-e^{-\mu \left(1/\sin\alpha +1/\sin\beta\right)d_{NNO}}}{\mu\left(1+\frac{\sin\alpha}{\sin\beta}\right)},
\end{equation}
where $\alpha$ and $\beta$ are the angles of incidence and emission, $\mu$ is the linear absorption coefficient and $d_{NNO}$ is the thickness of the NdNiO$_x$ layer. This is done because the absorption can vary strongly with $L$ for $L$ scans, impacting the peak position in $L$. The $L$ dependence of Sample D is, similar to sample B, not pronounced and changes in the peak intensity are comparable to that of the background shape; however the intensity appears to be at a maximum between 0.280 and 0.333 r.l.u., at roughly $\sim 3.1(2)$ r.l.u which is qualitatively similar to prior measurements, showing a weak peak around $L\sim\nicefrac{1}{3}$.

\subsection{Energy Dependence of the Resonant Feature}
\begin{figure}[ht]
  \resizebox{10.3 cm}{!}{
  \includegraphics{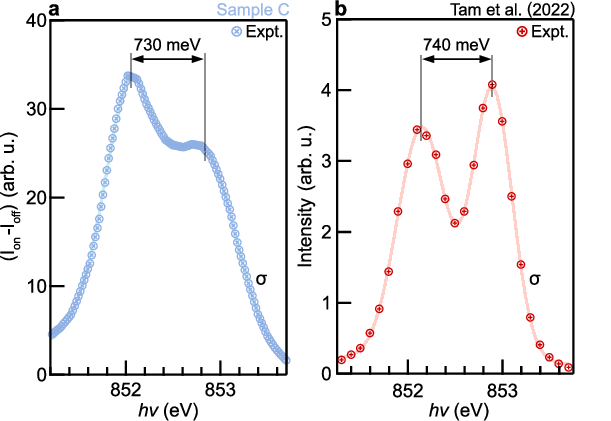}}
  \caption{\label{fig:resonant} Resonance profile comparison at the Ni $L_3$ edge. (a) Difference between the intensity at the scattering wavevector of $q=(0.333, 0, 0.290)$, $I_{\textrm{on}}$, and off the peak at $q=(0.310, 0, 0.300)$, $I_{\textrm{off}}$. (b) Data extracted and reproduced from Figure 1(e) of Tam et al. (2022) \cite{Tam2022}. Fitted peak positions are marked.}
\end{figure}
\begin{figure}[ht]
  \resizebox{15.1 cm}{!}{
  \includegraphics{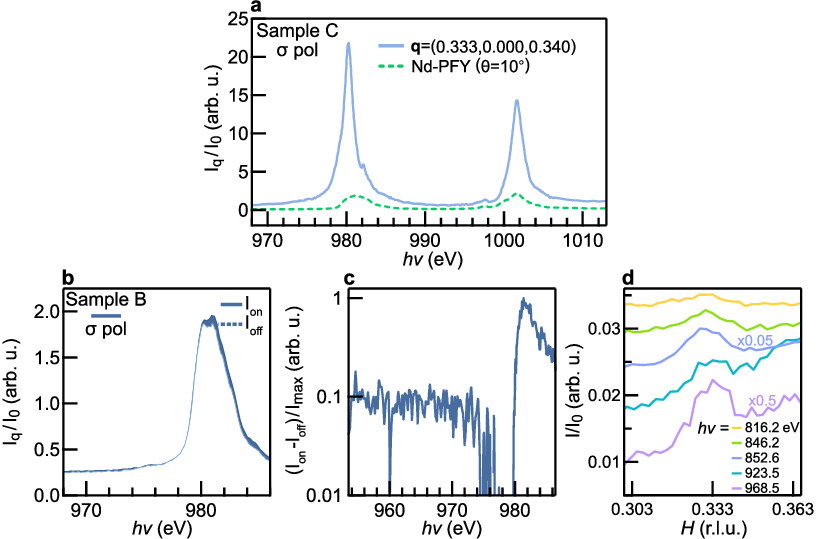}}
  \caption{\label{fig:energy2} Resonance profile comparison at the Nd $M_5$ edge. (a) Scattered intensity at fixed $\mathbf{q}=(0.333,0.000,0.340)$ as a function of photon energy, for sample C.  Grazing incidence ($\theta = 10^{\circ}$, $\sigma$~ polarization) Neodymium PFY measurement is included as an intensity reference. (b) Resonant energy profile of sample B at the Nd $M$ edge on the peak scattering wavevector, $I_{\textrm{on}}$, and off, $I_{\textrm{off}}$ (as a measure of the fluorescence background). (c) Difference between the intensity on and off the scattering wave vector, normalized to the maximum value. (d) Rocking curves on sample B, through $\mathbf{q}=(0.333,0.000,0.340)$, performed at different off-resonant energies; traces offset for clarity.}
\end{figure}

Additional data and a analysis of the resonant profiles of samples C and B are provided in \ref{fig:resonant} and \ref{fig:energy2}.  The resonance profile of the scattering peak (difference between the signal measured on the peak, and off -- to remove contributions from the background fluorescence) for (intermediate phase) Sample C is depicted in Figure \ref{fig:resonant}(a).  The two peak structure, with a splitting of 730 meV, matches well with prior measurements of the resonance profile by resonant inelastic x-ray scattering in Tam et al. \cite{Tam2022}.  The data from this reference is reproduced in \ref{fig:resonant}(b), where the similar peak splitting of 740 meV is apparent.  The similarity of these resonant profiles, including both the bifurcation and the overall energy width of the feature, suggests a common origin.  The resonance profile, at fixed $\mathbf{q}$, of Sample C at the Neodymium $M$ edge is pictured in \ref{fig:energy2} and shows strong enhancement of the scattering intensity at both the $M_4$ and $M_5$ peaks.  For reference, the fixed $\mathbf{q}$ resonance profile for the lightly reduced Sample B is shown in \ref{fig:energy2}(b-c) where a small, but measurable, enhancement above the fluorescence is observable.  Finally, rocking curves at a variety of photon energies well off of the Ni and Nd resonances for Sample B are provided in Figure \ref{fig:energy2}(d).  Just like samples C and D from the main text, the {\SLa} feature is weak, but observable well off resonance indicating a structural origin of this peak as well. 

\subsection{Additional Null Results}
\begin{figure}[ht]
  \resizebox{15.1 cm}{!}{
  \includegraphics{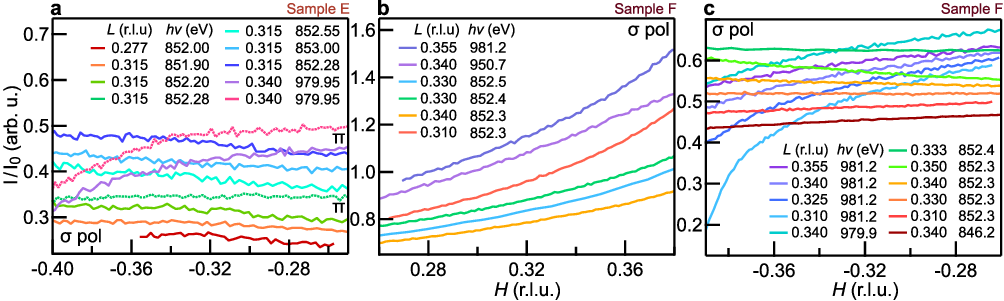}}
  \caption{\label{fig:clsnull} Addtitional RSXS measurements performed on samples E and F. Rocking curves centered at $(\pm\nicefrac{1}{3},0,L)$ at multiple values of $L$ and $h\nu$ show no indication of a peak at the nominal ordering wave-vector, {\SLa}, in either sample.  Traces have been offset, but not rescaled, for clarity.}
\end{figure}
Finally, in Figure \ref{fig:clsnull} we show additional null results on samples E and F from the main text.  Following careful alignment to the substrate bragg peaks a series of rocking curves was performed on either sample including measurements at and around both the Ni $L_3$ and Nd $M_5$ edges.  No peaks were observed at any energy or at any of the $L$ values tested in the vicinity of previous reports near {\SLb}.

\section{Characterization and measurement of samples G-J}

\subsection{Null Results on Additional Samples}
\begin{figure}
  \resizebox{16 cm}{!}{
  \includegraphics{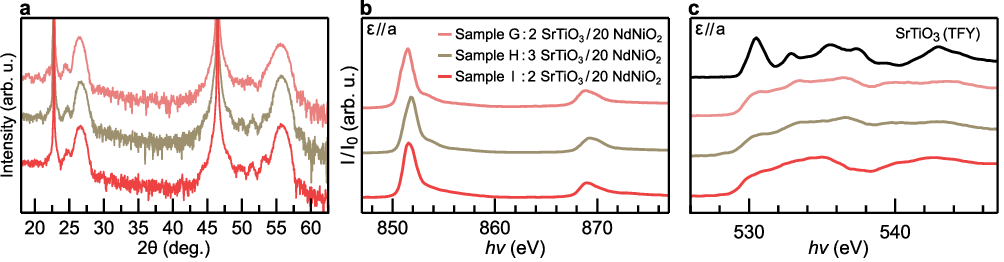}}
  \caption{\label{fig:alsXAS} Characterization of samples G, H, and I. (a) Lab based XRD measurements of all three samples showing the (001) and (002) difraction peaks. (b) Nickel $L$ edge XAS-TEY measurements showing the single peak strucure of NdNiO$_2$. (c) Oxygen $K$ edge XAS-TEY measurements showing no pre-peak at $\sim 528$~eV, a TFY measurement of a SrTiO$_3$ substrate is included for reference. Traces have been offset for clarity.}
\end{figure}

\begin{figure}
  \resizebox{17 cm}{!}{
  \includegraphics{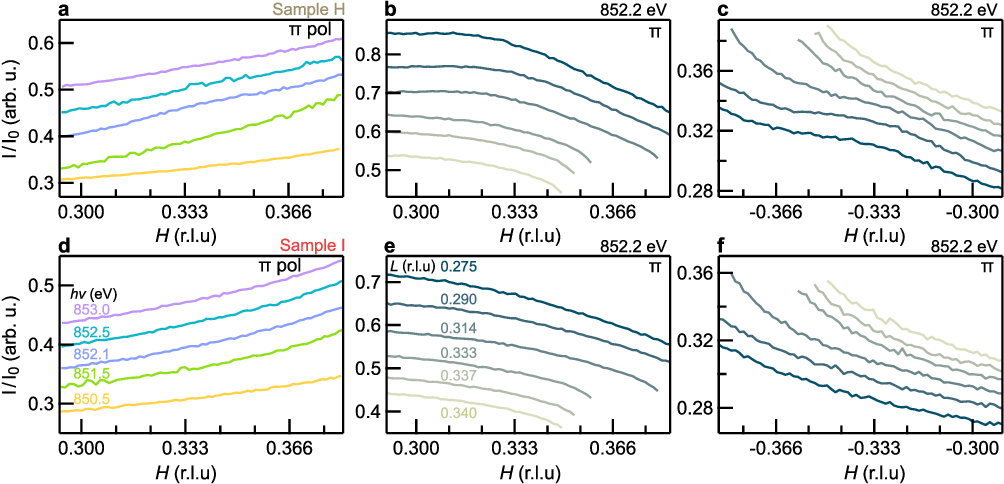}}
  \caption{\label{fig:alsNULL} Additional RSXS measurements performed on samples H and I at beamline 8.0.1 of the Advanced Light Source using a silicon photodiode detector.  (a) Rocking curves through $\mathbf{q}=(0.333,0,0.315)$ at different photon energies. (b-c) $H$ scans at constant $L$ at different values of $L$ ranging from 0.275 to 0.340 r.l.u about $H=\nicefrac{+1}{3}$ and $H=\nicefrac{-1}{3}$, respectively. (d-f) Same scans performed on sample I.}
\end{figure}

In this section we describe measurements on additional samples not featured in the main text, Samples G-I, as well as characterization details for the sample whose TEM profile is included in Figure 4 of the main text, Sample J.  Cu K-$\alpha$ XRD measurements of samples G, H, and I are shown in Figure \ref{fig:alsXAS}(a).  Samples G, H, and I were all prepared in nominally the same way as samples D-F in the main text with exception that sample G was exposed to the atomic hydrogen beam for a slightly lower time: 10 minutes, in contrast to the typical 12-15.  As a result, while the sample appears predominately infinite layer in composition, there are some weak shoulders visible decorating the (002) bragg peak at $2\theta =50.5^{\circ}$ and $53.5^{\circ}$, this is in contrast to samples H and I which show only the Pendell\"{o}sung fringe pattern of the main text samples.  XAS measurements at the Ni $L$ and O $K$ edges for the three samples are shown in Figure \ref{fig:alsXAS}(b-c) and have the expected single Ni $L_3$ peak at $\sim 852$ eV associated with NdNiO$_2$ and no visible O $K$ edge pre-peak associated with unreduced NdNiO$_3$.  Resonant scattering measurements on samples H and I are presented in Figure \ref{fig:alsNULL}, and spatially resolved soft and hard x-ray measurements on sample G are detailed in figure \ref{fig:microXRD}.  Sample H, shows weak signatures of a resonant feature at $H=\nicefrac{\pm 1}{3}$ shich is only barely visible in the rocking curve at $h\nu=852.1$ eV and not visible at most values of $L$ except between 0.275 and 0.314 r.l.u., as can be seen in panels (b-c) -- this peak has amplitude only 4\% above the background and FWHM of $>0.04$~ r.l.u ($\xi<10$~nm).  Sample I, shown in Figure \ref{fig:alsNULL} (d-f) shows no evidence of any resonant features at any values of $h\nu$ or $L$.

\subsection{Position Dependent RSXS and HXRD Measurements}
\begin{figure}[ht]
  \resizebox{16.4 cm}{!}{
  \includegraphics{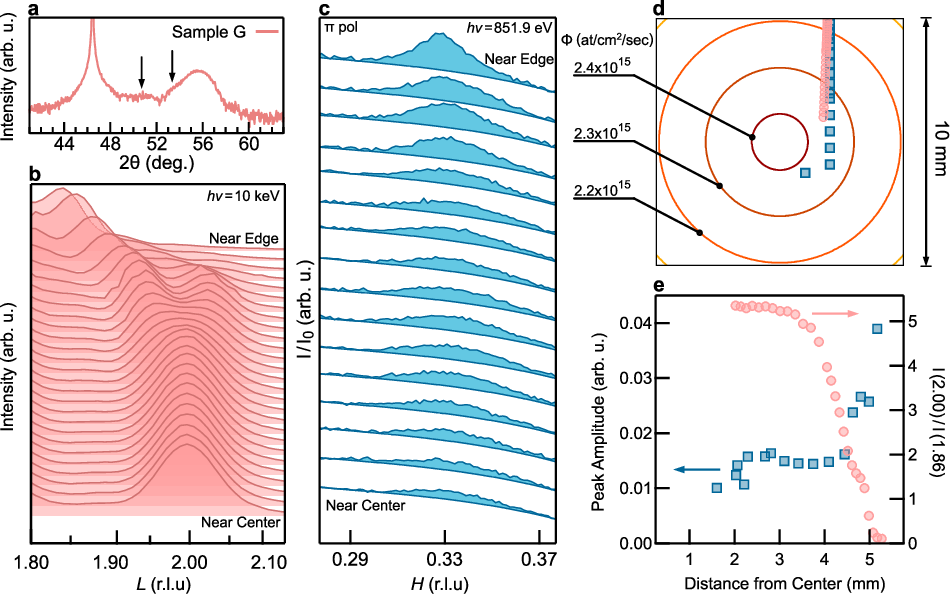}}
  \caption{\label{fig:microXRD} Spatially resolved hard and soft x-ray measurements of Sample G. (a) Lab based Cu K-$\alpha$ x-ray diffraction measurement with a beam width $>$ 10mm, covering the entire sample; positions of impurity peaks are indicated. (b) Sequence of hard x-ray diffraction (00L) scans taken at positions ranging from near the sample edge to near the sample center. (c) Position dependent soft x-ray rocking curves through $\mathbf{q}=\left(0.333,0,0.315\right)$ at the Ni $L_3$ resonance. (d) Diagram of a $10\times 10$~mm sample with the measurement positions from (b) and (c) marked. Estimated hydrogen fluxes at different positions on the sample are marked. (e) Resonant scattering peak amplitude and NdNiO$_2$ peak amplitude as a function of distance from the sample center.}
\end{figure}

Due to the expanding nature of the atomic hydrogen source, the flux at the sample surface is not perfectly uniform, and a drop of 15-20\% across the sample is expected based on characterization data of the source \cite{Tschersich1998,Tschersich2008}.  Additionally, as the sample is suspended at the corners, and heated radiatively from behind, there is typically a thermal gradient across the sample with the corners remaining slightly cooler then the center.  Both of these factors (reduced flux and sample temperature) are expected to cause the edges of the sample to reduce more slowly then the sample center.  To test this hypothesis, Sample G was reduced in a similar manner to the other NdNiO$_2$ samples reported in this study, but for a slightly shorter duration.  As a result some impurity peaks remain visible in area averaged XRD measurements of the (002) peak, indicated in Figure \ref{fig:microXRD}(a).  In Figure \ref{fig:microXRD}(b) micro-spot diffraction measurements about the (002) bragg peak are reported where the x-ray beam ($100~\mu\textrm{m}\times 100~\mu\textrm{m}$) is rastered from the sample edge towards the center. These spatially resolved measurements confirm that the sample edge has a clustering of intermediate phase peaks with lattice constants greater than NdNiO$_2$, even though the sample center appears well reduced to the infinite layer phase.  Later, this same sample was remeasured at low photon energies corresponding to the Ni $L_3$ resonance.  These spatially resolved RSXS measurements (Spot size: Vertical $\times$~ Horizontal $=30~\mu\textrm{m} \times 200~\mu\textrm{m}$), taken at positions coincident with the hard x-ray measurements, are detailed in panel (c) and the positions on the sample surface are indicated in (d).  As expected, the {\SLa} peak is strongest near the sample edge and tapers off approaching the center.  These results are summarized in Figure \ref{fig:microXRD}(e) where the fitted amplitude of the $(\nicefrac{1}{3},0)$ peak is compared to the intensity of the (002) bragg peak (normalized by the off-peak intensity).  The clear anticorrelation between the presence of the NdNiO$_2$ phase and the resonant feature strongly indicates that this feature is not intrinsic to the infinite layer phase.

\subsection{Characterization of Sample J}
\begin{figure}[ht]
  \resizebox{12 cm}{!}{
  \includegraphics{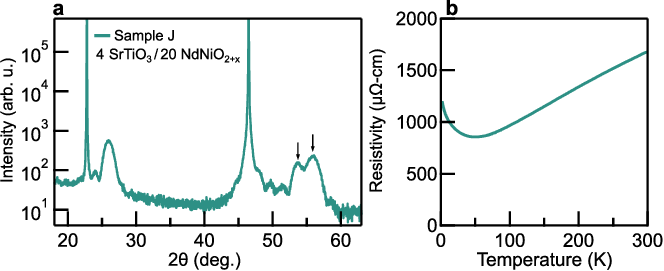}}
  \caption{\label{fig:sampleJ} Characterization of Sample J. (a) Lab-based Cu K-$\alpha$ X-ray diffraction. (b) Temperature dependent resistivity.}
\end{figure}
Basic characterization of Sample J is provided in Figure \ref{fig:sampleJ}.  Lab-based XRD measurements indicate that the sample is partially reduced, with a strong sharp peak occurring at the expected position of {\NNO}, $2\theta =55.86^{\circ}$.  An additional phase is clearly present in the film, as evidenced by the side peak at $2\theta =53.67^{\circ}$ (NB: This is nearly the same peak position of the shoulder observed in Sample G in Figure \ref{fig:microXRD}(a)).  The sample resistivity, shown in panel (b) is qualitatively similar to {\NNO} films presented here and in the literature, however the minimum resistivity, 860 $\mu\Omega$-cm, and low temperature resistivity, $\rho(2~\textrm{K})=1200\mu\Omega$-cm, are higher then observed in samples D, E, and F -- which is consistent with inclusions of insulating intermediate phases in the film.  The film was grown and reduced using the same conditions as D; however it possesses a thicker, 4 u.c. SrTiO$_3$, capping layer -- the presence of additional reduction products in the film is not unexpected, given that the conditions were not adjusted to account for the thicker capping layer.  
\bibliography{Nickelates}